\documentclass[11pt,epsfig]{elsart}

\usepackage{epsfig}
\usepackage{latexsym}
\usepackage{amsmath}
\usepackage{amsfonts}
\usepackage{graphics}
\usepackage{amssymb}

\begin{document}
\bibliography{book}
\def\figurewidth{16cm}
\bibliographystyle{unsrt}

\begin{frontmatter}



\title{Positronium-Atom Collisions}
\author{H.~R.~J.~Walters, A.~C.~H.~Yu, S.~Sahoo and Sharon Gilmore}
\address{Department of Applied
  Mathematics and Theoretical Physics,\\ Queen's University, Belfast,
  BT7 1NN, United Kingdom.}
\ead{h.walters@qub.ac.uk;
     a.yu@qub.ac.uk;
     s.sahoo@qub.ac.uk;
     s.gilmore@qub.ac.uk}
\begin{abstract}
New results are presented for Ps(1s) scattering by H(1s), He(1$^1$S) and Li(2s).
Calculations have been performed in a coupled state framework, usually employing
pseudostates, and allowing for excitation of both the Ps and the atom. In the
Ps(1s)-H(1s) calculations the H$^-$ formation channel has also been included
using a highly accurate H$^-$ wave function. Resonances resulting from unstable
states in which the positron orbits H$^-$ have been calculated and analysed. The
new Ps(1s)-He(1$^1$S) calculations still fail to resolve existing discrepancies
between theory and experiment at very low energies. The possible importance of
the Ps$^-$ formation channel in all three collision systems is discussed.
\end{abstract}
\end{frontmatter}
\section{Introduction}
In this article we present some new theoretical results for positronium(Ps)-atom
scattering for three fundamental systems: Ps-H, Ps-He and Ps-alkali metal,
taking Li as our example of an alkali metal. In each case we assume that the
incident Ps and the target atom are both in their ground states. Throughout we
use atomic units (au) in which $\hbar=m_e=e=1$, the symbol $a_0$ is used to
denote the Bohr radius, $a_0=\hbar^2/me^2$. To express energies derived
in atomic units, eg, the calculated position and width of a resonance, in
electron volts(eV) we have used the conversion 1 au=27.21138344(106) eV 
\cite{mohr00}. 
\section{Positronium Scattering by Atomic Hydrogen (Ps(1s)--H(1s))}
\subsection{Event Line}
\begin{figure}[h]
\centerline{\psfig{figure=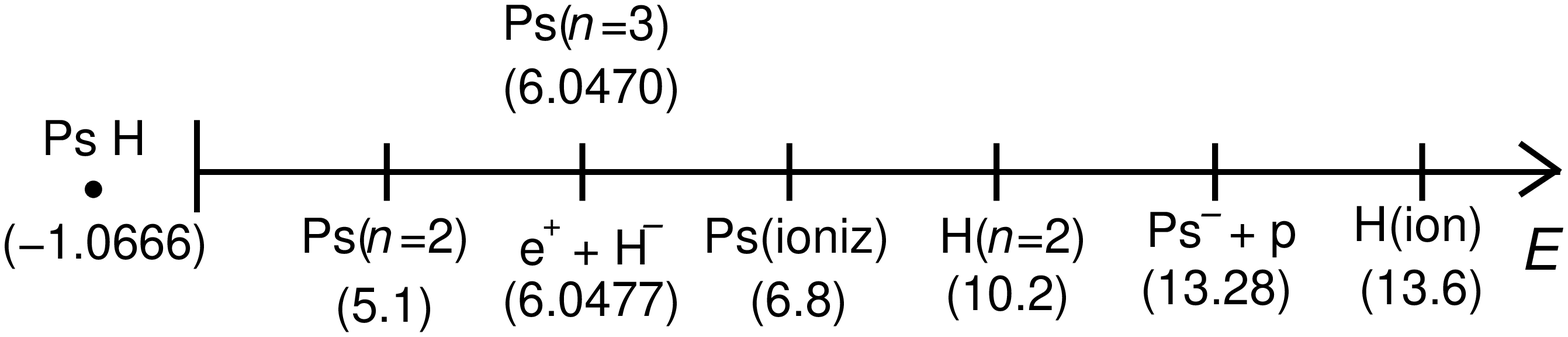,width=5.5in,height=2in}}
\caption{\small Event line for Ps(1s)-H(1s) scattering in the electronic spin 
singlet state. Events are shown as a function of the impact energy E(in eV). The
diagram is purely schematic and not to scale.}
\label{f1}
\end{figure}
Fig.~\ref{f1} shows an event line for this system. In a non-relativistic treatment of
Ps-H collisions the total electronic spin, $S_e$, is conserved. Since the system
contains two electrons the possible values of $S_e$ are 0 and 1. Fig.~\ref{f1}
illustrates the case of scattering in the electronic spin singlet state $S_e=0$.
From Fig.~\ref{f1} we see that the system possesses a single S-wave bound
state~\cite{ore51}, positronium hydride (PsH), with a binding energy of 1.0666 eV
\cite{yan99}. As far as scattering is concerned, the significance of this bound
state is that it corresponds mathematically to a pole at an impact energy of -1.0666 eV in the
S-wave singlet scattering amplitude. As we approach zero impact energy the
singlet elastic cross section, which becomes increasingly S-wave, rises towards
the pole, see Fig.~\ref{f2}. The precise value of the zero energy singlet cross
section in a calculation therefore depends upon how well this pole is represented.

Up until 5.1 eV only elastic Ps(1s)-H(1s) scattering is possible. At 5.1 eV
Ps(n=2) excitation becomes feasible, the atom still remaining in its ground
state. At 6.0470 eV Ps(n=3) excitation comes on line. Interestingly, this
threshold is almost coincident with that for ${\rm H}^-$ formation 
$({\rm Ps}+{\rm H} \Longrightarrow {\rm H}^-+{\rm e}^+)$ at 6.0477
eV~\cite{frolov94}. In quoting these numbers we have ignored relativistic effects
and have assumed that the proton has infinite mass. One wonders whether this
near degeneracy might present some interesting experimental opportunities.
Certainly, it should lead to a competition between the two channels and since,
as we shall see later (Fig.~\ref{f3}), there is a rich Rydberg resonance structure
associated with the ${\rm H}^-$ threshold, it presumably has some effect upon
this structure.

Between 6.0477 eV and 6.8 eV the full Rydberg spectrum of Ps becomes accessible
until, at last, at 6.8 eV the Ps can be ionized. Not until 10.2 eV can the atom
be excited, to the n=2 level with the Ps remaining in its ground state.
Ionization of the atom becomes possible at 13.6 eV. However, between 10.2 and
13.6 eV another interesting threshold appears, ${\rm Ps}^-$ formation 
$({\rm Ps}+{\rm H} \Longrightarrow {\rm Ps}^-+{\rm p})$, at 13.28
eV~\cite{frolov94}. This threshold lies in the midst of thresholds for producing
highly excited H (in fact between the H(n=6) and H(n=7) thresholds). As with
${\rm H}^-$ we would expect Rydberg resonance structure associated with the 
${\rm Ps}^-$ threshold, although probably much less pronounced.

Beyond the energy range of Fig.~\ref{f1}, double excitation of both projectile and
target becomes possible at 15.3 eV and double ionization at 20.4 eV.

Scattering in the electronic spin triplet $({\rm S}_e=1)$ channels is somewhat
less interesting, although usually the dominant component of any spin averaged
cross section. The event line for triplet scattering is the same as 
Fig.~\ref{f1} except that the PsH bound state and the ${\rm H}^-$ and ${\rm Ps}^-$ channels
are omitted. The absence of these formation channels means also a corresponding
absence of resonance structure.
\subsection{Coupled Pseudostate Calculations}\label{s3}
The challenge to theory is to represent the events portrayed in Fig.~\ref{f1} in
an adequate way. A suitable and powerful representation is provided by the
coupled pseudostate approach.

Let ${\bf r}_{\it p}({\bf r}_{\it i})$ be the position vector of the
positron(${\it i}$th electron) relative to the proton, which is assumed to be
infinitely massive. ${\bf R}_{\it i}\equiv({\bf r}_{\it p}+{\bf r}_i)/2\;$ and 
${\bf t}_{\it i}\equiv({\bf r}_{\it p}-{\bf r}_{\it i})\;$ then correspond to
the centre of mass of the Ps relative to the proton and to the Ps internal
coordinate when the Ps consists of the positron and the ${\it i}$th electron.
The non-relativistic Hamiltonian for the Ps-H system is then
\begin{eqnarray}\label{e1}
H=-\frac{1}{4}\nabla^2_{R_1}+H_{Ps}({\bf t}_1)+H_A({\bf r}_2)
+\frac{1}{r_p}-\frac{1}{r_1}-\frac{1}{|{\bf r}_p-{\bf r}_2|}
+\frac{1}{|{\bf r}_1-{\bf r}_2|}
\end{eqnarray}
where
\begin{equation}\label{e2}
H_{Ps}({\bf t})\equiv-\nabla^2_{t}-\frac{1}{t}
\end{equation}
is the Hamiltonian for Ps and
\begin{equation}\label{e3}
H_A({\bf r})\equiv-\frac{1}{2}\nabla^2_r-\frac{1}{r}
\end{equation}
is the Hamiltonian for atomic hydrogen. The Hamiltonian (\ref{e1}) is, of course,
unchanged by the interchange ${\bf r}_1\leftrightarrow{\bf r}_2$.

Following previous work~\cite{campbell98,blackwood02} we expand the collisional
wave function, $\Psi$, for the system in a particular state of total
electronic spin $S_e$ according to
\begin{eqnarray}\label{e4}
\Psi=\sum_{a,b}\biggl[G_{ab}({\bf R}_1)\phi_a({\bf t}_1)\psi_b({\bf r}_2)
+(-1)^{S_e}G_{ab}({\bf R}_2)\phi_a({\bf t}_2)\psi_b({\bf r}_1)\biggr]
\end{eqnarray}
where the sum is over Ps states $\phi_a$ and H states $\psi_b$. These states may
be either eigenstates or pseudostates and have the property that
\begin{equation*}
<\phi_a({\bf t})|H_{Ps}({\bf t})|\phi_{a^\prime}({\bf t})>=E_a{\delta_{aa^\prime}}\hspace*{0.5cm}
<\phi_a({\bf t})|\phi_{a^\prime}({\bf t})>=\delta_{aa^\prime}\nonumber
\end{equation*}
\begin{equation}\label{e5}
<\psi_b({\bf r})|H_{A}({\bf r})|\psi_{b^\prime}({\bf r})>=\varepsilon_b\delta_{bb^\prime\hspace*{0.5cm}}
<\psi_b({\bf r})|\psi_{b^\prime}({\bf r})>=\delta_{bb^\prime}
\end{equation}
Substituting (\ref{e4}) into the Schr$\ddot{\rm o}$dinger equation with the
Hamiltonian (\ref{e1}) and projecting with $\phi_a({\bf t}_1)\psi_b({\bf r}_2)$
leads to coupled equations for the functions ${\it G_{ab}}$ of the form
\begin{equation*}
(\nabla_{R_1}^2 + p_{ab}^2)G_{ab}({\bf R}_1)
= 4\sum_{a'b'}V_{ab,a'b'}({\bf R}_1)G_{a'b'}({\bf R}_1)\hspace*{3.5cm}\nonumber
\end{equation*}
\begin{equation}\label{e6}
\hspace*{3.8cm}+4(-1)^{S_e}\sum_{a'b'}\int L_{ab,a'b'}({\bf R}_1,{\bf
R}_2)G_{a'b'}({\bf R}_2)d{\bf R}_2
\end{equation}
In (\ref{e6}), $p_{ab}$ is the momentum of the Ps in the ab channel, $V_{ab,a'b'}$
gives the direct Coulombic interaction between the Ps and the H atom and
$L_{ab,a'b'}$ accounts for electron exchange between Ps and H.
The coupled equations (\ref{e6}) are converted to partial
wave form and solved using the R-matrix technique~\cite{burke75}.\\
\begin{table}[h]
\begin{center}
\caption{\small The 9 Ps and 9 H states of the 9Ps9H 
approximation of~\cite{blackwood02}. The energies are as defined in (\ref{e5}).}
\vspace{0.3cm}
\begin{tabular}{ccc}
\hline
&\multicolumn{2}{c}{Energy (eV)}\\  
\cline{2-3}
{State} & {Ps} & {H} \\
\hline
1{\it s} & 0.0 & 0.0 \\
2{\it s}, 2{\it p} & 5.1 & 10.2 \\
$\overline{3s}$, $\overline{3p}$, $\overline{3d}$ & 6.8 & 13.6 \\
$\overline{4d}$ & 13.2 & 26.5 \\
$\overline{4p}$ & 13.9 & 27.9 \\
$\overline{4s}$ & 17.1 & 34.2 \\
\hline
\end{tabular}
\label{t1}
\end{center}
\end{table}

The approximation (\ref{e4}) has been employed in~\cite{blackwood02} using 9 Ps
states and 9 H states, called the 9Ps9H approximation, and, for S-wave
scattering only, using 14 Ps and 14 H states (14Ps14H approximation). The 9
states are shown in table~\ref{t1}, a more detailed specification is given
in~\cite{blackwood02}. They consist of $1s, 2s, 2p$ eigenstates and
$\overline{3s}, \overline{3p}, \overline{3d}, \overline{4s}, \overline{4p}, 
\overline{4d}$ pseudostates (pseudostates are denoted by a ``bar"). The
pseudostates give a representation of the Ps/H continua as well as giving an
average approximation to the bound eigenstates with ${\rm n}\geq3$ (the n=3
pseudostates of table~\ref{t1} have approximately a 2/3 overlap with the 
${\rm n}\geq3$ eigenstate spectrum~\cite{kernoghan95}). While the 9Ps9H and
14Ps14H approximations should be satisfactory for describing electronic spin
triplet scattering, it has been shown in~\cite{blackwood02a}, but only within
the context of the frozen target approximation, that the ${\rm e}^+-{\rm H}^-$
channel exerts a profound effect upon singlet scattering. Here we try to improve
upon the earier calculations of~\cite{blackwood02} for singlet scattering by
explicitly adding on the ${\rm e}^+-{\rm H}^-$ channel, our
approximation~(\ref{e4}) for singlet scattering is then extended to 
\begin{eqnarray}\label{e7}
\hspace*{-1cm}\Psi=\sum_{a, b}\biggl[G_{ab}({\bf R}_1)\phi_a({\bf t}_1)\psi_b({\bf r}_2)
+(-1)^{S_e}G_{ab}({\bf R}_2)\phi_a({\bf t}_2)\psi_b({\bf r}_1)\biggr]
+F({\bf r}_p)\psi^-({\bf r}_2, {\bf r}_2)\nonumber\\
\end{eqnarray}
where $\psi^-$ is the ${\rm H}^-$ wave function. We refer to this as the 
$9{\rm Ps}9{\rm H} + {\rm H}^-$ or $14{\rm Ps}14{\rm H} + {\rm H}^-$
approximation, etc., depending upon the number of Ps and H states used in the
sum. For $\psi^-$ we have used a highly accurate 100 term Kinoshita-Koga type
wave function~\cite{kinoshita57,koga} giving an ${\rm H}^-$ binding energy of
0.0277510163 au. It should be noted that the calculation of~\cite{blackwood02a}
was restricted not only by the frozen target approximation but also by the use
of an approximate ${\rm H}^-$ wave function, these restrictions have now been
lifted in the present work.
\vspace{0.5cm}
\begin{table}[h]
\begin{center}
\caption{\small PsH bound state}
\vspace{0.3cm}
\begin{tabular}{cc}
\hline
\multicolumn{1}{c}{Approximation} & \multicolumn{1}{c}{Binding Energy (eV)} \\
\hline
9Ps1H & 0.543 \\
9Ps9H & 0.963 \\
14Ps14H & 0.994 \\
$9{\rm Ps}9{\rm H} + {\rm H}^-$ & 1.02 \\
$14{\rm Ps}14{\rm H} + {\rm H}^-$ & 1.03 \\
Accurate result of~\cite{yan99} & 1.0666 \\
\hline
\end{tabular}
\label{t2}
\end{center}
\end{table}

The first test of the approximation~(\ref{e7}) is how well it reproduces the PsH
binding energy. Table~\ref{t2} shows the present and earlier calculations
compared with the very accurate result of~\cite{yan99}. Here we see that the
frozen target approximation 9Ps1H gives only half of the binding energy. The
allowance for virtual target excitation in the 9Ps9H and 14Ps14H approximations
raises this to 90-93\%. Inclusion of the ${\rm e}^+-{\rm H}^-$ channel brings
further improvement, as we would expect, but our best result, 
$14{\rm Ps}14{\rm H} + {\rm H}^-$, still remains 3.4\% below the accurate value
of~\cite{yan99}. We estimate that our numerical methods are good enough to yield
an answer correct to better than 1\%. The deviation of 3.4\% from the accurate
result of~\cite{yan99} is therefore significant. What is missing? The
approximation~(\ref{e7}) should represent satisfactorily correlation in which
one electron is associated primarily with the proton and the other electron with
the positron (the sum in~(\ref{e7})) and in which the two electrons are strongly
correlated in association with the proton while the positron moves more freely
around this complex (the $\psi^-$ term). What is absent is correlation in which
the two electrons and the positron are associated in a strongly correlated unit
moving in the field of the proton, in short, a ${\rm Ps}^-$ structure orbiting the
proton. We speculate that the addition of the ${\rm Ps}^- + {\rm p}$ channel
to~(\ref{e7}) might lead to significant improvement.

\begin{table}[h]
\begin{center}
\caption{\small S-wave phase shifts (in radians) and scattering length for
electronic spin singlet Ps(1s)-H(1s) scattering.}
\vspace{0.3cm}
\begin{tabular}{cccc}
\hline
\multicolumn{1}{c}{Incident} &&& \multicolumn{1}{c}{Van Reeth and}\\
\multicolumn{1}{c}{Momentum (a.u.)} & \multicolumn{1}{c}{14Ps14H}
& \multicolumn{1}{c}{$14{\rm Ps}14{\rm H}+{\rm H}^-$} 
& \multicolumn{1}{c}{Humberston~\cite{vanreeth03}} \\ 
\hline
0.1 & -0.434 & -0.428 & -0.425 \\
0.2 & -0.834 & -0.825 & -0.817 \\
0.3 & -1.178 & -1.167 & -1.158 \\
0.4 & -1.467 & -1.453 & -1.443 \\
0.5 & -1.704 & -1.685 & -1.674 \\
0.6 & -1.890 & -1.867 & -1.852 \\
0.7 & -2.018 & -1.992 & -1.959 \\
Scattering Length (a.u.) & 4.41 & 4.327 & 4.311 \\
\hline
\end{tabular}
\label{t3}
\end{center}
\end{table}

Recent accurate low energy S-wave calculations by Van Reeth and
Humberston~\cite{vanreeth03} using the Kohn variational principle present a
further opportunity to test our approximations. In table~\ref{t3} we make a
comparison of our best calculations for the S-wave singlet phase shifts and
scattering length with these accurate numbers. We see that inclusion of the
${\rm e}^++{\rm H}^-$ channel roughly halves the difference between the 14Ps14H
approximation and the variational results for the phase shifts. With the
exception of the last point at 0.7 au, our best phase shifts, in the
$14{\rm Ps}14{\rm H}+{\rm H}^-$ approximation, now differ from the variational
numbers by about 1\%. The agreement between the $14{\rm Ps}14{\rm H}+{\rm H}^-$
scattering length and the variational answer is particularly good.

The largest calculation that we have made for higher partial waves is in the 
$9{\rm Ps}9{\rm H}+{\rm H}^-$ approximation. In table~\ref{t4} we list the phase
shifts in this approximation for S-, P-, D-wave scattering.
Comparing tables~\ref{t3} and~\ref{t4}, we see that the S-wave phase shifts in
the $9{\rm Ps}9{\rm H}+{\rm H}^-$ approximation are only marginally better than
those in the $14{\rm Ps}14{\rm H}$ approximation which does not include the 
${\rm e}^+-{\rm H}^-$ channel.
\vspace{0.5cm}
\begin{table}[h]
\begin{center}
\caption{\small Electronic spin singlet S-, P-, and D-wave phase shifts (in
radians) in the $9{\rm Ps}9{\rm H}+{\rm H}^-$ approximation. Powers of 10 are
denoted in parentheses.}
\vspace{0.3cm}
\begin{tabular}{cccc}
\hline
\multicolumn{1}{c}{Incident} &\multicolumn{1}{c}{}&\multicolumn{1}{c}{}
&\multicolumn{1}{c}{}\\
\multicolumn{1}{c}{Momentum (a.u.)} & \multicolumn{1}{c}{S}
& \multicolumn{1}{c}{P} & \multicolumn{1}{c}{D} \\ 
\hline
0.1 & -0.432 & 0.221(-1) & 0.202(-3) \\
0.2 & -0.833 & 0.183 & 0.349(-2) \\
0.3 & -1.179 & 0.580 & 0.173(-1) \\
0.4 & -1.466 & 0.956 & 0.522(-1) \\
0.5 & -1.699 & 1.106 & 0.116 \\
0.6 & -1.884 & 1.134 & 0.208 \\
0.7 & -2.012 & 1.133 & 0.324 \\
\hline
\end{tabular}
\label{t4}
\end{center}
\end{table}

Fig.~\ref{f2} shows S, P, and D electronic spin singlet elastic partial wave
cross sections in the energy range 0 to 3.5 eV. The inadequacy of the frozen
target approximation, 9Ps1H, in this energy range is clear. The large zero
energy 9Ps1H S-wave cross section is the result of the small PsH binding energy
obtained in this approximation, see table~\ref{t2}. Consequently, the PsH bound
state pole in the scattering amplitude is much closer to zero impact energy than
it should be, see Fig.~\ref{f1}, with the result that the cross section rises to
too high a value at zero energy. Fig.~\ref{f2} also illustrates the change in the
9Ps9H cross sections on including the ${\rm e}^+-{\rm H}^-$ channel.

\begin{figure}
\centerline{\psfig{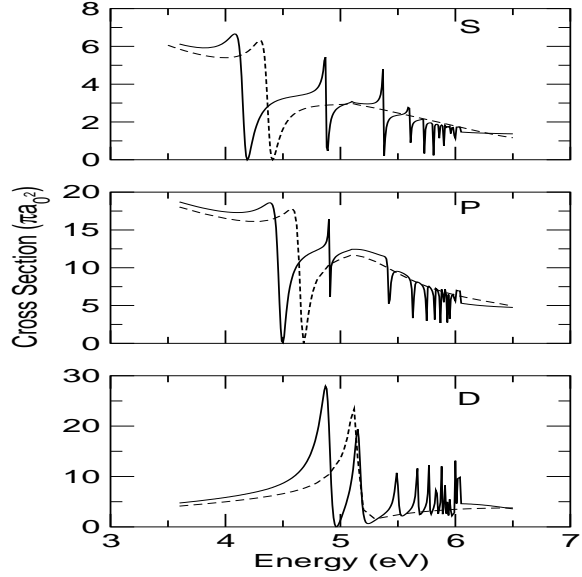}}
\caption{\small Electronic spin singlet partial wave cross sections for
Ps(1s)-H(1s) elastic scattering in the energy range 3.5 to 6.5 eV.
Approximations: solid curve, $9{\rm Ps}9{\rm H}+{\rm H}^-$; dashed curve, 9Ps9H.}
\label{f3}
\end{figure}

Fig.~\ref{f3} shows the same cross sections in the energy range 3.5 to 6.5 eV.
Here we see pronounced resonance structure associated with unstable states of
the positron trapped in the field of the ${\rm H}^-$
ion~\cite{drachman79,blackwood02a}. The 9Ps9H approximation only gives the
first of these resonances, and at too high an energy. To see the profound impact
of the full resonance structure one needs the $9{\rm Ps}9{\rm H}+{\rm H}^-$
approximaton, i.e., one needs to include the ${\rm e}^++{\rm H}^-$ channel
explicitly in the approximation.

We have fitted the positions and widths of the first few resonances, these are
given in table~\ref{t5} where comparison is made with the complex coordinate
rotation results of Yan and Ho~\cite{yan99,yan98,ho98,ho00}. We see that the
first member of each partial wave series in the $9{\rm Ps}9{\rm H}+{\rm H}^-$
approximation lies higher in position than the complex coordinate prediction. By
contrast, the second member lies lower. Generally speaking, the agreement on
positions and widths between the two theoretical results leaves something to be 
desired.
\vspace{0.5cm}
\begin{figure}[!ht]
\centerline{\psfig{figure=SPD_graph.eps,width=3in,height=3in}}
\caption{\small Electronic spin singlet partial wave cross sections for
Ps(1s)-H(1s) elastic scattering in the energy range 0 to 3.5 eV. Approximations:
solid curve, $9{\rm Ps}9{\rm H}+{\rm H}^-$; dashed curve, 9Ps9H; dash-dot curve,
9Ps1H.}
\label{f2}
\end{figure}
\begin{table}[pht]
\begin{center}
\caption{\small Positions and widths (in parentheses) for electronic spin
singlet S-, P-, D- and F-wave resonances (in eV).}
\vspace{0.3cm}
\begin{tabular}{ccc}
\hline
\multicolumn{1}{c}{} & \multicolumn{1}{c}{$9{\rm Ps}9{\rm H}+{\rm H}^-$}
& \multicolumn{1}{c}{Yan and Ho}\\
\multicolumn{1}{c}{Resonance} & \multicolumn{1}{c}{Approximation}
& \multicolumn{1}{c}{\cite{yan99,yan98,ho98,ho00}}\\  
\hline
S(1) & 4.149 & 4.0058 $\pm$ 0.0005 \\
     & (0.103) & (0.0952 $\pm$ 0.0011) \\
\hline
S(2) & 4.877 & 4.9479 $\pm$ 0.0014 \\
     & (0.0164) & (0.0585 $\pm$ 0.0027) \\
\hline
S(3) & 5.377 & 5.3757 $\pm$ 0.0054 \\
     & (0.0091) & (0.0435 $\pm$ 0.011) \\     
\hline
P(1) & 4.475 & 4.2850 $\pm$ 0.0014 \\
     & (0.0827) & (0.0435 $\pm$ 0.0027) \\ 
\hline
P(2) & 4.905 & 5.0540 $\pm$ 0.0027 \\
     & (0.0043) & (0.0585 $\pm$ 0.0054) \\ 
\hline
D(1) & 4.899 & 4.710 $\pm$ 0.0027 \\
     & (0.0872) & (0.0925 $\pm$ 0.0054) \\
\hline
D(2) & 5.161 &  \\
     & (0.0648) &  \\     
\hline
D(3) & 5.496 &  \\
     & (0.0328) &  \\
\hline
F(1) & 5.200 & 5.1661 $\pm$ 0.0014 \\
     & (0.0095) & (0.0174 $\pm$ 0.0027) \\     
\hline
F(2) & 5.494 &  \\
     & (0.0262) &  \\        
\hline
F(3) & 5.661 &  \\
     & (0.0294) &  \\        
\hline 
\end{tabular}
\label{t5}
\end{center}
\end{table}

\begin{figure}[pht]
\centerline{\psfig{figure=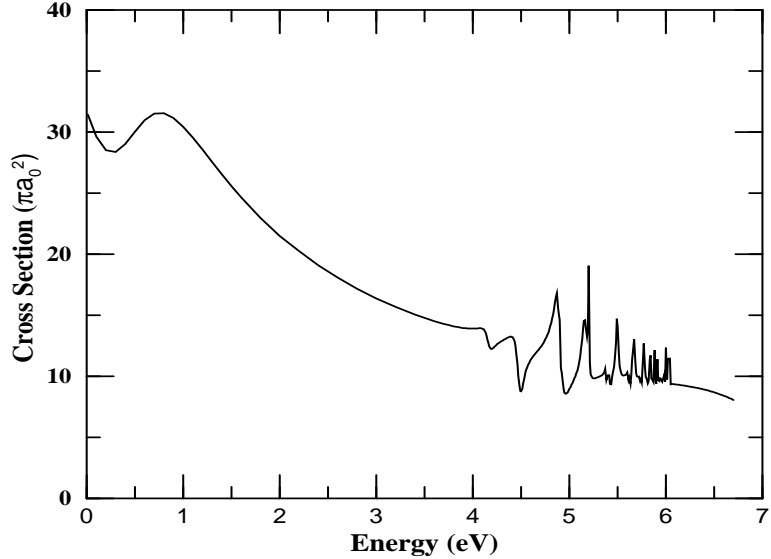,width=4in,height=3in}}
\caption{\small Total cross section for Ps(1s)--H(1s) scattering.}
\label{f5}
\end{figure}

By combining the electronic spin triplet results in the 9Ps9H approximation
from~\cite{blackwood02} with the present electronic spin singlet results in the 
$9{\rm Ps}9{\rm H}+{\rm H}^-$ approximation we have calculated the
Ps(1s)+H(1s) total cross section in the energy range 0 to 6.7 eV, fig.~\ref{f5}.
This cross section shows the spectacular effect of the singlet resonance
structures. It should be noted that the cross section of figure~\ref{f5} 
assumes that the target H atom is spin unpolarized and that no spin
analysis is made of the final states~\cite{blackwood02,blackwood02a}; the cross
section is independent of whether the incident Ps(1s) is ortho or para and, if
ortho, of its spin polarization. Partial waves with total angular momentum J
from 0 to 4 have been used in calculating the cross section of figure~\ref{f5}.
\subsection{Conclusions}
It is probably fair to say that we now have a pretty good overall idea of
Ps(1s)-H(1s) scattering in the energy range up to 6.7 eV. However, within the
context of the coupled pseudostate approach some details remain to be cleared
up---the PsH binding energy, the exact details of the resonances, better
convergence towards the Kohn variational phase shifts. We speculate that these
may be resolved by explicit inclusion of the ${\rm Ps}^-+{\rm p}$ channel and,
for the resonances, by taking account of the near degeneracy of the 
${\rm e}^+-{\rm H}^-$ and Ps(n=3) channels which all of the approximations used
here fail to do (the Ps(n=3) states used here are pseudostates rather than
eigenstates, see table~\ref{t1}). In addition it would be interesting to see
what resonance structure might be associated with the ${\rm Ps}^-+{\rm p}$
channel. These are matters for future investigation. 
\section{Positronium Scattering by Helium (Ps(1s)--He(1$^1$S))}
\begin{figure}[pht]
\centerline{\psfig{figure=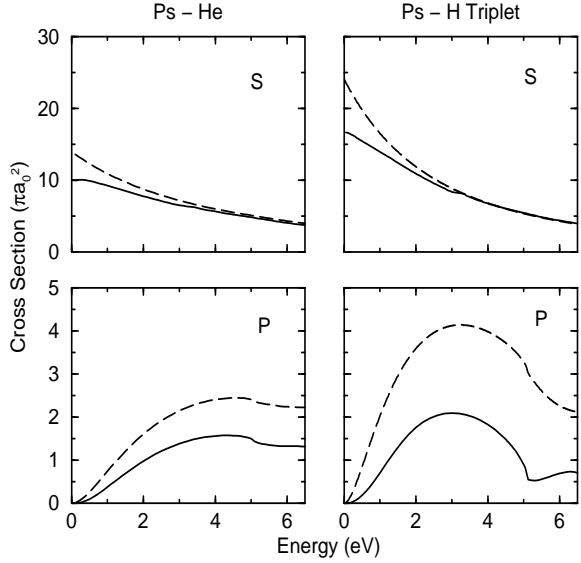,width=3in,height=3in}}
\caption{\small S- and P- wave cross sections for Ps(1s) elastic scattering by
He(1$^1$S) and H(1s). The H cross sections are for scattering in the 
electronic spin triplet state. For He: solid curve, 9Ps9He approximation for
S-wave and 7Ps7He approximation for P-wave; dashed curve, 9Ps1He approximation
for S-wave and 7Ps1He approximation for P-wave. For H: solid curve, 9Ps9H
approximation; dashed curve, 9Ps1H approximation.}
\label{f6}
\end{figure}

In a large frozen target calculation Blackwood et al~\cite{blackwood99} have highlighted
significant discrepancies between theory and theory, experiment and experiment, and theory
and experiment for very low energy o-Ps(1s)--He(1$^1$S) scattering. The work on
atomic hydrogen~\cite{blackwood02}~(see figures~\ref{f2} and ~\ref{f6})
illustrates nicely the deficiencies of the frozen target approximation at low
energies and the need to allow for virtual target excitation. Here we report
some new coupled pseudostate calculations for He which relax the frozen target
assumption. The generalization of~(\ref{e4}) to the case of Ps(1s)--He(1$^1$S)
scattering is
\begin{eqnarray}\label{e8}
\Psi=A\sum_{S_A=0,1}\sum_{a,b}\sum_mC(1/2,S_A,1/2,m,-m,0)\nonumber\\
\times G^{S_A}_{ab}({\rm\bf R}_1)\phi_a({\rm\bf t}_1)\psi^{S_A}_{b}({\rm\bf r}_2,
{\rm\bf r}_3)\zeta_{m}({\rm s}_1)\chi^{S_A}_{-m}({\rm s}_2,{\rm s}_3)\hspace*{-0.5cm}
\end{eqnarray}
where the electron space and spin coordinates are now (${\rm\bf r}_i, s_i)$ 
$(i=1, 2, 3)$, $\phi_a$ is the Ps state, $\zeta_m(s)$ the spin function for the Ps
electron with Z-component $m$, $\psi^{S_A}_{b}$ is the spatial part of a He
state with total electronic spin $S_A$ (=0,1), and $\chi^{S_A}_{m}$ is the
corresponding spin function. Since the total electronic spin of the 
Ps(1s)--He(1$^1$S) system is 1/2, the Ps and He states need to be coupled
together to give total spin 1/2, that is the purpose of the Clebsch-Gordan
Coefficient $C$ in~(\ref{e8}). Finally, $A$ is the electron antisymmetrization
operator.

We make two approximations on~(\ref{e8}). The first is to neglect the He triplet
states $S_A=1$. The second is to approximate the He singlet states in the form
\begin{eqnarray}\label{e9}
\psi^{S_A=0}_b({\rm\bf r}_2, {\rm r}_3)=N_b(\psi_b({\rm\bf
r}_2)\bar{\psi}({\rm r}_3)
+\psi_b({\rm\bf r}_3)\bar{\psi}({\rm r}_2))
\end{eqnarray}
where $\bar{\psi}({\rm r})$ is taken to be the He$^+$(1s) orbital 
$\sqrt{8/\pi}e^{-2r}$ and $N_b$ is a normalisation constant. The orbitals
$\psi_b({\rm\bf r})$ are at our disposal to choose as we see appropriate. The
form~(\ref{e9}) is the same as that used in the e$^{+}-$He scattering
calculations of~\cite{campbell98a}. Using~(\ref{e9}) we have constructed 9He
states analogous to the 9H
states of table~\ref{t1} and labelled as 1$^1$S, 2$^1$S, 2$^1$P, 
$\overline{3^1{\rm S}}$, $\overline{3^1{\rm P}}$, $\overline{3^1{\rm D}}$,    
$\overline{4^1{\rm S}}$, $\overline{4^1{\rm P}}$, $\overline{4^1{\rm D}}$. As in
table~\ref{t1}, the n=3 states are constructed so as to sit at the
ionization threshold of He(1$^1$S) at 24.58 eV. The 1$^1$S, 2$^1$S and 2$^1$P
states are, of course, now just approximations to the He(1$^1$S), He(2$^1$S) and
He(2$^1$P) eigenstates.

Combining the 9 He states described above with the 9 Ps states from
table~\ref{t1} (a 9Ps9He approximation analogous to 9Ps9H) we have calculated an
S-wave Ps(1s)--He(1$^1$S) elastic scattering cross section in the energy range 0
to 6.5 eV, figure~\ref{f6}. Because we encountered some bad numerical behaviour
with the higher partial waves, we have evaluated the P-wave cross section in a
reduced approximation, 7Ps7He, in which the Ps and He d-states have been
dropped. The P-wave cross section is shown in figure~\ref{f6}. Also included in
figure~\ref{f6} are the corresponding frozen target cross sections calculated in
the 9Ps1He (S-wave) and 7Ps1He (P-wave) approximations. It has been remarked
that Ps(1s)--He(1$^1$S) scattering should be similar to Ps(1s)--H(1s) scattering
in the electronic spin triplet state since in both cases antisymmetry keeps the
positronium electron and the atomic electron(s) apart, consequently we have 
added the triplet 9Ps9H and 9Ps1H S- and P-wave cross sections to 
figure~\ref{f6} for comparison. We see a similar pattern for both H and He which
gives us some confidence in our calculations. In each case allowance for virtual
target excitation produces a noticeable reduction on the frozen target results. 
For He this effect seems to be smaller than for H, presumably on account of the 
higher excitation energies for a He target.

Since the original calculations of Blackwood et al~\cite{blackwood99} three 
other relevant pieces of work have appeared in the literature. In~\cite{basu01}
Basu et al have included both Ps and He excitations in a 
Ps(1s, 2p)+He(1$^1$S, 2$^1$S, 2$^1$P) coupled eigenstate calculation. They
obtain a zero energy elastic cross section of 7.40 $\pi a_0^2$. This contrasts
with our zero energy cross section of 9.9 $\pi a_0^2$ in the much larger 9Ps9He
coupled pseudostate calculation. In~\cite{mitroy02} Mitroy and Ivanov have used a model
potential approximation which yields values for the zero energy cross section
ranging from 10.6 $\pi a_0^2$ to 8.8 $\pi a_0^2$ depending upon the choice of
parameters in the potential. Interestingly, their average cross section of 9.83 
$\pi a_0^2$ agrees well with our present results. Finally, Chiesa et
al~\cite{chiesa02} have employed the diffusion Monte Carlo method to calculate 
S-wave phase shifts in the momentum range 0 to 0.4 au. They get a zero energy 
cross section of 7.89 $\pi a_0^2$.
\begin{figure}[pht]
\centerline{\psfig{figure=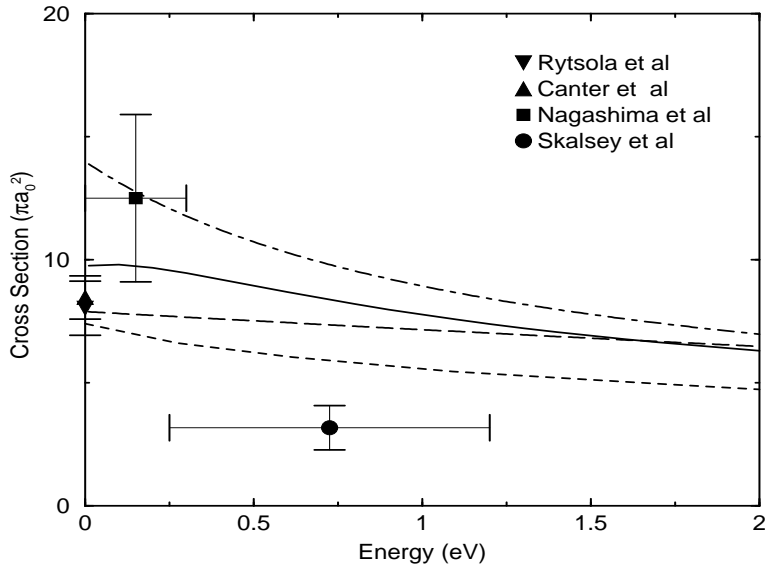,width=4in,height=3in}}
\caption{\small Momentum transfer cross sections for Ps(1s)+He(1$^1$S)
scattering. Theory: solid curve, present results with 9Ps9He approximation for
S-wave and 7Ps7He approximation for P-wave; dash-dot curve, frozen target
results with 9Ps1He approximation for S-wave and 7Ps1He approximation for
P-wave; long-dashed curve, S-wave total cross section of Chiesa et
al~\cite{chiesa02}; short-dashed curve, cross section of Basu et
al~\cite{basu01}. Experiment: square, Nagashima et al~\cite{nagashima98}; up
triangle, Canter et al~\cite{canter75}; down triangle, Ryts{\"o}l{\"a} et 
al~\cite{ryts84}; circle, Skalsey et al~\cite{skalsey03}.}
\label{f7}
\end{figure}

In figure~\ref{f7} we make a comparison with the low energy experimental data.
As discussed in~\cite{blackwood02b}, these low energy data, derived from annihilation
measurements, correspond to the momentum transfer cross section.
\begin{equation}\label{e10}
\sigma_{mom}=\int\,(1-{\rm cos}\,\theta)\,\frac{d\,\sigma_{el}}{d\,\Omega}d\,\Omega
\end{equation}
where $d\,\sigma_{el}/d\,\Omega$ is the elastic differential cross section and
$\theta$ is the scattering angle. At zero impact energy where the scattering is
all S-wave $\sigma_{mom}$ is identical with the S-wave total cross section.
However, as pointed out in~\cite{blackwood02b}, $\sigma_{mom}$ can diverge 
rapidly from the total cross section with increasing energy. In figure~\ref{f7} 
the theoretical results correspond to the momentum transfer cross section. The 
only exception is the cross section of Chiesa et al~\cite{chiesa02} where we 
have only S-wave data and so the cross section shown is just the S-wave total 
cross section. The cross section of Basu et al has been constructed out of the
phase shift data given in their paper~\cite{basu01}. Figure~\ref{f7} shows that
our frozen target approximation agrees well with the experimental point of
Nagashima et al~\cite{nagashima98}. Our present results are also in agreement
with this measurement and close to, but outside, the error bars of the cross
sections of Canter et al~\cite{canter75} and Ryts{\"o}l{\"a} et
al~\cite{ryts84}. Most striking, however, is the agreement between these two
measurements and the S-wave total cross section of the sophisticated Monte Carlo
calculation of Chiesa et al which, as indicated above, coincides with the
momentum transfer cross section at zero energy. Another striking point is the
lack of agreement between any of the theories and the cross section of Skalsey
et al~\cite{skalsey03} centred on 0.725 eV. At this energy our calculations and
that of Basu et al show that $\sigma_{mom}$ is about 20\% smaller than the total
cross section, indicating the importance of P-wave scattering. To get agreement
with the measurement of Skalsey et al would require a more spectacular growth in
P-wave scattering.

There are a number of ways in which the present calculations can be improved.
Firstly, there is a need to eliminate any doubts concerning the use of an
approximation to the He ground state wave function, see (\ref{e9}). Secondly, there is
the question of the importance of He triplet states in the expansion~(\ref{e8}).
Finally, we have remarked upon the similarity of Ps(1s)--He(1$^1$S) scattering
and Ps(1s)--H(1s) triplet scattering, see figure~\ref{f6}, but there is one
mechanism in Ps(1s)--He(1$^1$S) scattering which is not available to the 
Ps(1s)--H(1s) triplet case, that mechanism is Ps$^-$ formation
(Ps(1s)+He(1$^1$S)$\Longrightarrow$Ps$^-$+He$^+$(1s))~\cite{walters01}. This might be
a more important mechanism than has been realised and may, perhaps, resolve the
discrepancy between theory and the experiment of Skalsey et al~\cite{skalsey03}.
In addition, this mechanism would be a possible source of resonances in 
Ps(1s)--He(1$^1$S) scattering although the resonance structure may be small in
magnitude.
\section{Positronium Scattering by Lithium ( Ps(1s)--Li(2s))}
\begin{figure}[pht]
\centerline{\psfig{figure=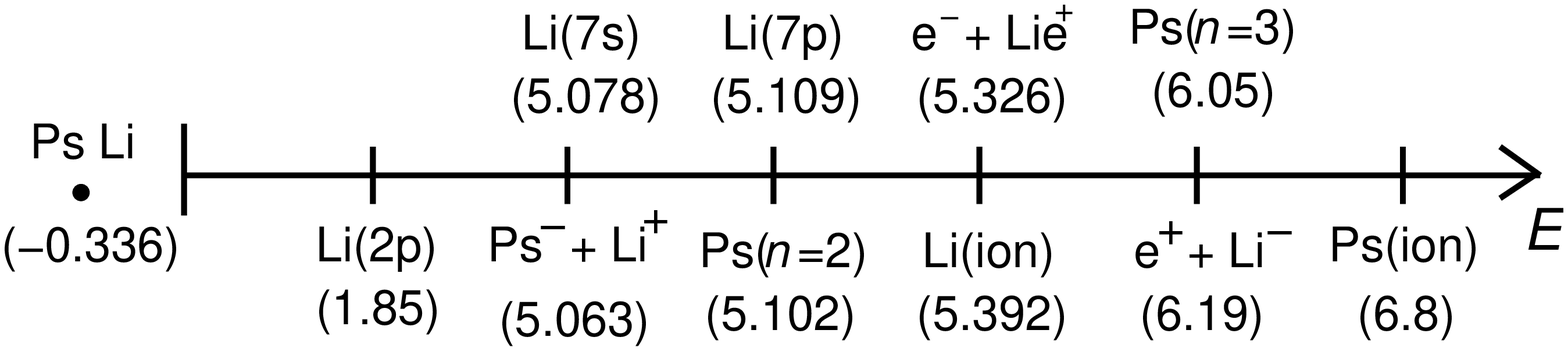,width=5.5in,height=2in}}
\caption{\small Event line for Ps(1s)--Li(2s) scattering in the electronic spin
singlet state. Events are shown as a function of the impact energy E (in eV).
The diagram is purely schematic and not to scale. For scattering in the
electronic spin triplet state, the PsLi bound state and the Ps$^-$+Li$^+$ and
e$^+$+Li$^-$ formation channels should be omitted.}
\label{f8}
\end{figure}

Alkali metals behave in many respects like a one-electron system in which the
single valence electron revolves outside a frozen core. A priori, one might
therefore think that Ps-alkali scattering would be similar to Ps--H scattering.
That this is not so is suggested by the event line for Ps(1s)--Li(2s) scattering
shown in figure~\ref{f8}. Compared with figure~\ref{f1} for Ps(1s)--H(1s),
everything is ``reversed". Yes, there is a bound state of Ps and
Li~\cite{mitroy01} analogous to PsH, but with increasing energy it is atom
excitation and ionization that preceed Ps excitation and ionization, the 
opposite of figure~\ref{f1}. We also see that the order of Ps$^-$ and Li$^-$ 
formation is reversed compared to that of Ps$^-$ and H$^-$ formation shown in
figure~\ref{f1}. Ps$^-$ formation presumably therefore plays a much more
prominent role for the alkali systems than for H, and, in particular, with
regard to resonance formation.

An interesting feature which figure~\ref{f1} does not possess is a channel
corresponding to the formation of a bound state of the positron with the atom,
in figure~\ref{f8} the e$^-$+Lie$^+$ channel. The binding energy of positronic
lithium, as Lie$^+$ is called, is very small and the threshold for this channel
(at 5.326 eV) is almost degenerate with that for ionization of the Li atom (at
5.392 eV), lying only 0.066 eV below~\cite{ryzhikh98}. Again, a priori, one
would expect Rydberg resonance structure associated with this channel, but of
what amplitude is hard to predict. Unlike the Ps$^-$+Li$^+$ and e$^+$+Li$^-$
channels which also promise Rydberg resonance structure but only in electronic
spin singlet scattering, the positronic lithium channel couples to both
electronic spin singlet and electronic spin triplet scattering.

We are unable at this point to execute an approximation which does justice to
figure~\ref{f8}. Rather, we have elected for a simple coupled eigenstate
approximation to get a rough feeling of how things might go. Earlier coupled
eigenstate calculations have been published by Ray~\cite{ray99},
Biswas~\cite{biswas00}, and Chakraborty et al~\cite{chakraborty02}, but in all
cases either the Ps or the Li atom has been frozen in its ground state. Here we
publish the first calculations which allow for excitation of both the Ps and the
Li atom.

Our approximation is Ps(1s, 2s, 2p)+Li(2s, 2p). We have constructed the Li(2s)
and Li(2p) valence orbitals using the model potential of Stein~\cite{stein93}.
The collision formulation is the same as that given in equations~(\ref{e1}) to
(\ref{e6}) of section~\ref{s3} for Ps--H, except that we also include potentials
V$_p$(r$_p$) and -V$_e$(r$_i$) to allow for the interaction of the positron and
the electrons with the frozen 1s$^2$ core of the Li atom~\cite{yu03}. Our
approximation contains a representation of the van der Waals interaction,
-C$_6$/R$^6$, with C$_6$=288 au. In a much larger calculation of C$_6$ using
pseudostates for the Ps but keeping just the Li(2s) and Li(2p) states, we get 
C$_6$=451 au. The present coupled eigenstate calculation therefore includes
about 60\% of the full van der Waals effect. For Ps(1s)--H(1s), C$_6$=34.8
au~\cite{blackwood02}. The van der Waals force is therefore an order of
magnitude stronger for Ps interacting with Li. Our approximation gives a PsLi
bound state with binding energy 0.224 eV, somewhat less than 0.336 eV, the most
accurate calculated value~\cite{mitroy01}.
\begin{figure}[pht]
\centerline{\psfig{figure=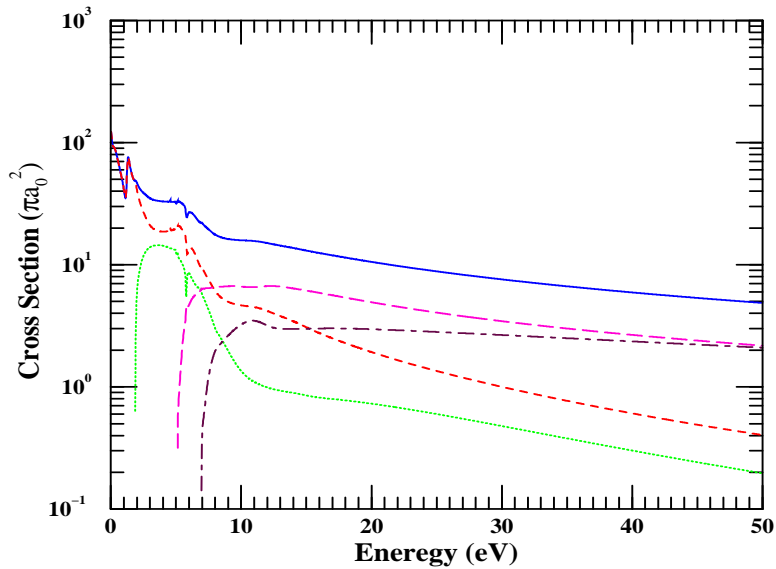,width=4in,height=3in}}
\caption{\small Cross sections for Ps(1s) scattering by Li(2s) as calculated in
the Ps(1s, 2s, 2p)+Li(2s, 2p) approximation. Curves: solid, total cross section;
short-dashed, elastic; dotted, Ps(1s)+Li(2p); long-dashed, Ps(n=2)+Li(2s); 
dash-dot, Ps(n=2)+Li(2p).}
\label{f9}
\end{figure}

Figure~\ref{f9} shows our calculated total cross section and its components. We
have assumed that the Li atom is spin unpolarized and that final state spins are
not resolved, the cross section is then 1/4 times singlet plus 3/4 times
triplet~\cite{blackwood02}. Our results apply either for o--Ps or p--Ps 
scattering and are independent of any spin polarization of the 
o--Ps~\cite{blackwood02}. From figure~\ref{f9} we see that at very low energies
we can expect an elastic cross section of the order of 100 $\pi a_0^2$, the spin
singlet component of this cross section is much larger as a result of the small
binding energy of the PsLi pole in the singlet scattering amplitude. To get the
zero energy cross section right is going to require an approximation with an
accurate representation of the PsLi binding. Away from threshold the elastic
cross section falls rapidly but then, with the opening of the Li(2p) channel at
1.85 eV (figure~\ref{f8}), it experiences a sudden rise resulting in a pronounced
structure. The Ps(1s)+Li(2s)$\Longrightarrow$Ps(1s)+Li(2p) cross section increases
rapidly from threshold, reaches a maximum of 17 $\pi a_0^2$ at 2.1 eV, then
quickly turns over and, comparatively speaking, is negligible beyond 10 eV.
Beyond 8 eV the Ps(1s)+Li(2s)$\Longrightarrow$Ps(n=2)+Li(2s) cross section is
dominant. The opening of this channel leads to structure in the other open
channels near 5.1 eV. By 50 eV the double excitation cross section
Ps(1s)+Li(2s)$\Longrightarrow$Ps(n=2)+Li(2p) has risen to meet the 
Ps(n=2)+Li(2s) curve. The importance of double excitation at high energies is 
not surprising~\cite{mcalinden96}. Our model predicts that Ps(n=2) excitation,
irrespective of the final state of the Li atom, will be dominant at high
energies. However, we should not accept that, in reality, this will be so. We
suspect that, in a more realistic treatment allowing for ionization of the Ps,
it will, as in other 
cases~\cite{campbell98,blackwood02,blackwood99,blackwood02b}, be Ps
ionization which is dominant at high energies with Ps(n=2) discrete excitation
being much less significant. The problem is that, in the present model, there is
nowhere for Ps excitation to go but into the Ps(n=2) channels. Ps excitation
which otherwise would flow into ionization is possibly being deflected into the
Ps(n=2) channels. Clearly a more detailed calculation is required.
\section{Conclusions}
We have presented new coupled state calculations for Ps(1s) scattering by H(1s) 
in the electronic spin singlet state, by He(1$^1$S) at very low energies, and by
Li(2s). One interesting theme that emerges from all three cases is the question
of the importance of Ps$^-$ formation, whether real or virtual. For Ps(1s)-H(1s)
scattering we speculate that the inclusion of virtual Ps$^-$ formation may be
necessary to tune the calculations into agreement with the accurate variational
results of Van Reeth and Humberston~\cite{vanreeth03} and with accurate bound
state calculations of PsH binding~\cite{yan99}. For Ps(1s)-He(1$^1$S)
scattering, virtual Ps$^-$ formation in the reaction 
Ps(1s)+He(1$^1$S)$\Longrightarrow$Ps$^-$+He$^+$(1s) provides us with a mechanism
for breaking away from the pattern of Ps(1s)-H(1s) triplet scattering which
present coupled state calculations on He seem to follow, and perhaps a route to
agreement with the seemingly anomalous experimental result of Skalsey et
al~\cite{skalsey03}, figure~\ref{f7}. For Ps(1s) scattering by Li(2s), or indeed
any alkali, we have, because of the energetics, the very interesting possibility
that Ps$^-$ formation, both real and virtual, may play a much more profound role
than had ever been envisaged. Clearly the study of Ps$^-$ formation is an
important direction for future research.

For Ps(1s)-H(1s) scattering another interesting feature which so far has not
been properly treated, is the near degeneracy of the Ps(n=3) excitation channels
and the H$^-$ formation channel, this must surely have some significant effect
on the resonance structure associated with the H$^-$ threshold. For
Ps(1s)-He(1$^1$S) scattering there is a need to know about sensitivity to the
use of approximate He(1$^1$S) target wave functions, as well as the role of He
triplet states in the collisional wave function expansion~(\ref{e8}). For
Ps(1s)--alkali scattering most of the really interesting physics would seem to
lie at impact energies below 10 eV, which is a challenge to experimentalists.
The challenge to theory in this case is to represent the Ps$^-$ and alkali
negative ion channels, and to a lesser extent the positronic lithium channel, as
well as the Ps and alkali atom channels. Without an adequate description of the 
ion channels it is unlikely that an accurate calculation of resonances can be 
made~\cite{biswas00}. Also, because of the small binding energies of Ps-alkali 
bound states~\cite{mitroy01} and their role as poles in the scattering 
amplitude, near threshold electronic spin singlet cross sections for 
Ps(1s)-alkali scattering will be very large and very sensitive to the pole 
position. Consequently, any realistic treatment of Ps(1s)-alkali scattering at 
low energies will need to incorporate a reasonably accurate representation of 
the Ps-alkali bound state.
\section{Acknowledgements}
This research was supported by EPSRC grants GR/N07424 and GR/R83118/01. We are
also greatly indebted to Prof. T. Koga for supplying us with the accurate H$^-$
wave function. 






%

\end{document}